\documentclass[useAMS,usenatbib,usegraphicx]{mn2e}
\usepackage[usenames,dvipsnames,svgnames,hyperref,amsmath]{xcolor}
\usepackage{soul}
\defcitealias{srisawat13}{S13}
\usepackage[
    plainpages=true,
    pdfpagelabels,
    bookmarks=true,
    bookmarksopen=false,
    bookmarksopenlevel=2
    bookmarksnumbered=true,
    bookmarkstype=toc,
    colorlinks=true,
    citecolor=RoyalBlue,
    linkcolor=ForestGreen,
    menucolor=Teal,
    urlcolor=DarkOrange,
]{hyperref}

\newcommand{\consistenttree}{\textsc {Consistent Trees}}
\newcommand{\dtree}{\textsc{D-Trees}}
\newcommand{\mergertree}{\textsc{MergerTree}}
\newcommand{\gadget}{\textsc{Gadget-3}}
\newcommand{\hbt}{\textsc{HBT}}
\newcommand{\jmerger}{\textsc{JMerge}}
\newcommand{\subfind}{\textsc{Subfind}}
\newcommand{\sublink}{\textsc{SubLink}}
\newcommand{\treemaker}{\textsc{TreeMaker}}
\newcommand{\velociraptor}{\textsc{VELOCIraptor}}
\newcommand{\ysamtm}{\textsc{ySAMtm}}

\begin{document}

\title[Halo Merger Trees in SAM]{Sussing Merger Trees : The Impact of Halo Merger Trees on Galaxy Properties in a Semi-Analytic Model}
\author[J.\ Lee et al.]{Jaehyun Lee,$^{1}$ Sukyoung K.\ Yi,$^{1}$\thanks{E-mail: yi@yonsei.ac.kr} Pascal J.\ Elahi,$^{2}$ Peter A.\ Thomas,$^{3}$ 
 \newauthor Frazer R.\ Pearce,$^{4}$ Peter Behroozi,$^{5,6}$ Jiaxin Han,$^{7}$ John Helly,$^{7}$ Intae Jung,$^{1,8}$
  \newauthor  Alexander Knebe,$^{9}$ Yao-Yuan Mao,$^{5,10}$ Julian Onions,$^{4}$ Vicente Rodriguez-Gomez,$^{11}$ 
   \newauthor  Aurel Schneider,$^{3}$  Chaichalit Srisawat,$^{3}$ and Dylan Tweed$^{12}$ \\
$^1$Department of Astronomy and Yonsei University Observatory, Yonsei University, Seoul 120-749, Republic of Korea\\
$^2$Sydney Institute for Astronomy, University of Sydney, Sydney NSW 2016, Australia\\
$^3$Department of Physics and Astronomy, University of Sussex, Brighton BN1 9QH, UK\\
$^4$School of Physics and Astronomy, University of Nottingham, Nottingham NG7 2RD, UK\\
$^5$Kavli Institute for Particle Astrophysics and Cosmology and Physics Department, Stanford University, Stanford, CA 94305, USA\\
$^6$Space Telescope Science Institute, Baltimore, MD 21218, USA\\
$^{7}$Institute for Computational Cosmology, Department of Physics, Durham University, South Road, Durham DH1 3LE, UK\\
$^{8}$Department of Astronomy, The University of Texas at Austin, Austin, TX, 78712, USA\\
$^{9}$Departamento de F\'{i}sica Te\'{o}rica, M\'{o}dulo C-15, Facultad de Ciencias, Universidad Aut\'{o}noma de Madrid, \\E-28049 Cantoblanco, Madrid, Spain\\
$^{10}$SLAC National Accelerator Laboratory, Menlo Park, CA 94025, USA\\
$^{11}$Harvard-Smithsonian Center for Astrophysics, 60 Garden Street, Cambridge MA, 02138, USA\\
$^{12}$Center for Astronomy and Astrophysics, Shanghai Jiao Tong University, Shanghai 200240, China}

\maketitle

\begin{abstract}
A halo merger tree forms the essential backbone of a semi-analytic model for galaxy formation and evolution. Recent studies have pointed out that extracting merger trees from numerical simulations of structure formation is non-trivial; different tree building algorithms can give differing merger histories. These differences should be carefully understood before merger trees are used as input for models of galaxy formation. We investigate the impact of different halo merger trees on a semi-analytic model. We find that the $z=0$ galaxy properties in our model show differences between trees when using a common parameter set. The star formation history of the Universe and the properties of satellite galaxies can show marked differences between trees with different construction methods. Independently calibrating the semi-analytic model for each tree can reduce the discrepancies between the $z=0$ global galaxy properties, at the cost of {\em increasing} the differences in the evolutionary histories of galaxies. 
Furthermore, the underlying physics implied can vary, resulting in key quantities such as the supernova feedback efficiency differing by factors of 2. Such a change alters the regimes where star formation is primarily suppressed by supernovae. Therefore, halo merger trees extracted from a common halo catalogue using different, but reliable, algorithms {\em can} result in a difference in the semi-analytic model. Given the uncertainties in galaxy formation physics, however, these differences may not necessarily be viewed as significant.
\end{abstract}


\section{Introduction}
Dark matter haloes play a crucial role in galaxy formation and evolution. They provide gravitational potential wells that are deep enough to gather baryons together within a Hubble time~\citep{white78, efstathiou83, blumenthal84}. Stars are born within cold gas clouds which either form at the centre of dark haloes via cooling~\citep{cowie77} or within filaments of cold material which pervade the large scale structure of the Universe~\citep{keres05, ocvirk08, keres09a, keres09b, ceverino10}. These stars form and feed galaxies which are assembled into galaxy groups or clusters via the hierarchical clustering of their haloes. Eventually, halo mergers give rise to galaxy mergers, as subhaloes fall into the central region of their main halo. Thus, the evolution of haloes is directly involved in the formation and evolution of galaxies.

One of the leading approaches for modeling galaxies in the $\Lambda$CDM cosmology uses Semi-Analytic Models of galaxy formation (SAM). This approach uses the evolution of dark matter haloes to ``paint'' galaxies using phenomenological prescriptions of the baryonic physics governing galaxy formation. Two different methodologies are used to produce the halo merger trees that are the backbones of SAMs. Halo merger trees can be built by taking advantage of the extended Press-Schechter formalism~\citep{bond91} and Monte Carlo simulations. Many SAMs have utilised this analytic method because it allows for the rapid construction of merger trees in large volumes with high mass resolutions~\citep{somerville99,cole00,khochfar05,somerville08,benson10,ricciardelli10}.~\citet{jiang14} made a comparison between several algorithms developed to construct halo merger trees using the extended Press-Schechter formalism. They showed that even when using the same formalism, differences in the details of the algorithms make halo merger trees with different growth properties. Halo merger trees can also be constructed directly from N-Body simulations. Despite the time needed to run N-body simulations with high-resolution and the complicated process of finding haloes, this method has become popular because it results in a more realistic evolutionary history of the haloes in various environments \citep{kauffmann99,hatton03,delucia04,croton06,bower06,guo11,lee13}.  

A series of processes are required to convert raw N-body simulation data into a format applicable to semi-analytic models. First, haloes are identified from the sea of hundreds of millions of particles in an N-body simulation.~\citet{knebe11,knebe13} demonstrated that most of the widely-used halo-finding codes generate similar results; however, poor resolution or dense environments can be problematic in terms of identifying substructures~\citep[e.g.][]{muldrew11, onions13, elahi13}. These halo-finding codes are then used to create halo catalogues, from which halo merger trees are constructed.~\citet[][hereafter S13]{srisawat13} analyzed the diversity of halo merger trees constructed by different tree building algorithms from a common halo catalogue. They stipulated the necessary conditions for reliable tree building codes: the codes should be able to trace particle transfers in order to link haloes in sequential snapshots; analyze more than two snapshots simultaneously to restore any missing subhaloes, especially in dense regions; and repair any transient artifacts in the halo properties, such as a sharp mass fluctuation. The influence of the specific halo finder on merger trees has been studied by \citet{avila14} who concluded that although the choice of halo finding algorithm itself was at least as important as the tree building routine the same types of issues arose for all the available halo finding methods. 

Many previous studies using semi-analytic models have focused predominantly on improving the phenomenological prescriptions of the baryonic physics governing galaxy formation as the vast uncertainties in these processes are critical in understanding the observable Universe. Furthermore, aforementioned studies dedicated to ``the dark universe" imply that technical issues concerning dark matter halo merger trees can also impose other uncertainties when modeling the observable universe. Thus, this study was proposed at the ``Sussing Merger Trees'' workshop, as one of the follow-up studies for \citetalias{srisawat13}, to examine the effect of different merger tree building algorithms on galaxy properties using a semi-analytic model.

This paper is organized as follows. In Section 2, we briefly introduce the features of halo merger tree algorithms. In Section 3, we describe the semi-analytic model used in this study. Then, we demonstrate the impact of halo merger trees on galaxies in the semi-analytic model in Section 4, discussing the differences in galaxy properties and looking into whether different trees can produce similar results by calibrating the various parameters. In Section 5, we summarise and present our conclusions.

\section{Algorithms for building dark matter halo merger trees}

The halo catalogue used in this study was extracted from a cosmological N-body volume simulation of structure formation run using \gadget~\citep{springel05c} with the cosmological parameters derived from the {\it Wilkinson Microwave Anisotropy Probe} seven year observations~\citep{komatsu11}, $\Omega_{\rm m}=0.272$, $\Omega_{\Lambda}=0.728$, and $h=0.704$. The periodic cube size of the volume is $62.5h^{-1} {\rm Mpc}$ on a side and the volume contains $270^3$ particles, resulting in a particle mass resolution of $9.31 \times 10^{8} h^{-1} M_{\odot}$. To produce halo merger trees, 61 snapshots distributed between $z=50$ and $0$ were used. After friends-of-friends (FOF) groups were identified in each snapshot, we used \subfind~\citep{springel01} to find sub-structures in the FOF groups. For this work we set a mass threshold of $\sim 2\times10^{10} h^{-1} M_{\odot}$ (20 particles) in order to limit ourselves to haloes that can be reliably identified.

\begin{figure}
\centering 
\includegraphics[width=0.45\textwidth]{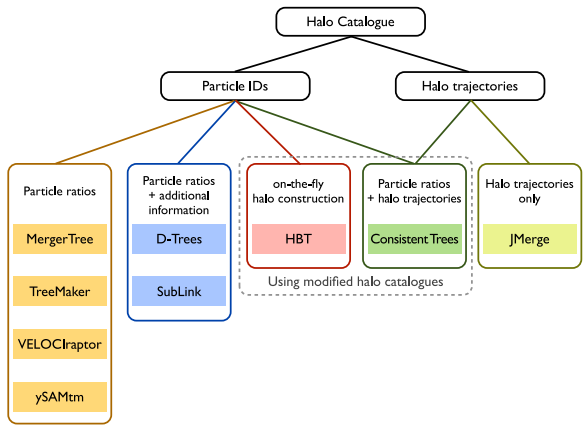}

\caption{Main features of merger tree building algorithms described in \citetalias{srisawat13}}
\label{algorithm}
\end{figure}

In this study, we used halo merger trees that were generated using various tree building algorithms that have been independently developed by nine groups. All of the trees were constructed from the common halo catalogue, save in two special cases. Figure~\ref{algorithm} briefly shows the main features of the algorithms. The codes grow their trees principally using particle identifiers (IDs) in each halo and/or halo trajectory information, such as the positions and velocities of the haloes. \mergertree~\citep{knebe10, srisawat13}, \treemaker~\citep{tweed09}, \velociraptor~\citep{elahi11,srisawat13}, and \ysamtm~\citep{jung14} investigate particle transfer between haloes in two snapshots, one taken after the other, and then determine the most probable direct descendants. This scheme, however, may be insufficient to build sound merger trees. If a halo is close to a more massive halo or embedded in a dense environment, it could be misidentified or even missed by the halo finder. Thus, in order to insert the missing links into the halo merger trees, \dtree~\citep{jiang13} and \sublink\ (Rodriguez-Gomez et al., in prep.) utilize additional information. These processes evaluate how tightly particles are bound to haloes to identify the most likely descendants. Furthermore, they analyze the particle ratios in more than two consecutive snapshots.

Unlike other algorithms, the Hierarchical Bound Tracing (\hbt) algorithm~\citep{han12} builds halo merger trees while it finds subhaloes. Starting from an input main halo catalogue, \hbt\ builds merger trees of main haloes by matching particle IDs between adjacent snapshots. Once a merger is identified, the progenitor haloes are tracked in subsequent snapshots, and the self-bound remnant of progenitor haloes are identified as descendent subhaloes. As a result, every subhalo identified by \hbt\ must have an explicit progenitor that traces back before infall, with no missing link along its evolution history. Because \hbt\ identifies its own set of subhaloes, it does not build merger trees from external subhalo catalogues. For the purpose of this study, we apply \hbt\ to the main haloes in the commonly supplied halo catalogue, and allow \hbt\ to generate a merger tree together with its own list of subhaloes.


\consistenttree~\citep{behroozi13} is the only method that makes use of both particle IDs and halo trajectories. It first constructs primary merger trees by examining the origin of particles in descendant haloes, similar to the aforementioned algorithms. However, it additionally calculates the motion of the haloes using position, velocity, and mass profile information returned by the halo finder. By comparing the primary merger trees with these motion calculations, \consistenttree\ updates its original halo catalogue with the halo properties calculated by the additional processes. Thus, \consistenttree\ also produces a modified halo catalogue, like \hbt.

\jmerger~\citep{srisawat13} uses halo trajectories only to identify the descendant-progenitor relationship between two snapshots. It calculates the expected forward and backward positions of haloes in snapshot $N$ and snapshot $N+1$ at the midway point between the two outputs. With imposed limits on the allowed change in position, mass and maximum circular velocity, \jmerger\ matches haloes between the two snapshots. Thus, it is sensitive to time resolution. Further details of the aforementioned algorithms are described in \citetalias{srisawat13}.

\section{Semi-analytic model}

In this study, we used the semi-analytic model developed by~\citet{lee13}. This model has simple, but appropriate, ingredients to deal with the motion of subhaloes and baryonic physics for galaxy formation and evolution.

\subsection{Merger trees for semi-analytic model}

\begin{table}
 \centering
  \caption{Number and fraction of haloes. By column, the halo finding algorithm, the total number of haloes surviving to $z=0$ after removing those with abnormal histories, the number of main haloes and the number of subhaloes. These are the haloes that survived the tree cleaning process of ySAM (see text).}

  \begin{tabular}{cccc}
  \hline
   Algorithm &  $N_{\rm halo}$ & $N_{\rm main}$  & $N_{\rm sub}$\\
 \hline
\consistenttree\	& 48999 & 32975 (67.3\%)   & 16024 (32.7\%) \\ 
\dtree\			& 65233 &  33794 (51.8\%)  & 31439 (48.2\%) \\
\hbt\ 			& 72388 &  32041 (44.3\%) & 40347 (55.7\%)\\
\jmerger\ 		& 41750 &  31168 (74.7\%) & 10582 (25.3\%) \\
\mergertree\ 		& 54539 & 32891 (60.3\%)  & 21648 (39.7\%)\\
\sublink\ 		& 64907 &  33485 (51.6\%) & 31422 (48.4\%)\\
\treemaker\  		& 54573 &  32874 (60.2\%) & 21699 (39.8\%)\\
\velociraptor\ 	        & 54546 &  32891 (60.3\%) & 21655 (39.7\%)\\
\ysamtm\ 		& 54871 &  32902 (60.0\%) & 21969 (40.0\%)\\
 
\hline
\end{tabular}
\label{abnormal}
\end{table}

We modified the merger trees that were constructed using the algorithms described in \citetalias{srisawat13} so that they would be applicable to our semi-analytic model. We first identified all the branches that corresponded to abnormal haloes: main haloes that disappear suddenly without merging and subhaloes that appear suddenly within a main halo without a physical trace in the previous snapshots. We then remove these from the merger trees. Such ``abnormal" haloes happen when halo finders have difficulty identifying haloes from limited numbers of particles or when tree building algorithms do not correctly trace the descendants or progenitors between snapshots. If an algorithm fails to make a link between two haloes in two sequential snapshots, a halo in a later snapshot appears to be a new halo, losing its previous growth history. 

Table 1 shows the number of haloes supplied to our semi-analytic model after removing abnormal branches. The total number varies from tree to tree ranging from 41750 to 72988. While the number of main haloes, $N_{\rm main}$, are similar to each other, the number of objects finally identified as subhaloes, $N_{\rm sub}$, is remarkably different between the algorithms. Most of the small difference in $N_{\rm main}$ between algorithms is attributed to the process in our SAM that detects and removes the abnormal main haloes that suddenly appeared at $z=0$ without progenitors. They are fortunately rare, and thus correction for them makes little difference in $N_{\rm main}$.

The number of subhaloes varies between the algorithms. \dtree, \hbt, and \sublink, each of which has additional processes to minimize  missing links between snapshots, have more subhaloes than the others. For example, \hbt\ has four times more subhaloes than \jmerger. \consistenttree\ was developed to minimize missing haloes but presents the second least number of subhaloes after \jmerger. This is due to the fact that \consistenttree\ removes subhalo branches with big mass fluctuations, suspected to be the results of misidentification by halo finders. The difference in $N_{\rm sub}$ leads to a difference in halo and galaxy merger rates and histories, which in return affects galaxy properties.

The real issue of missing haloes is that the tree building algorithms with higher missing halo fractions will have younger haloes as these missing haloes can re-appear at some point and they would be considered to be newborn haloes. As a result, these haloes will contain galaxies that are artificially young. Fortunately, the fraction of the newborn haloes that are in actuality descendants of the missing haloes is only of order 1\% when we inspect haloes of $M_{\rm 200} > 5.96\times10^{10}h^{-1}M_{\odot}$. This criterion corresponds to galaxy's stellar mass $M_{*} > 2\times10^8M_{\odot}$ at $z=0$ according to the stellar-to-halo mass relation of~\citet{moster10}. The missing halo fraction naturally increases substantially as we lower the halo mass cut, as it is more difficult to identify smaller haloes and measure their properties accurately.

\subsubsection{Tracking missing subhaloes}
As demonstrated above some subhaloes suddenly appear without having a prior history as independent haloes. Most of these are found in actuality to be the descendants of haloes missed by the halo finder in the previous snapshot, often due to the dense, crowded environment. Thus for this study we remove any branches of the tree that are due to subhaloes that are suddenly born.

For the same argument, a subhalo can disappear before reaching 0.1$R_{\rm 200}$ of their main haloes. In this case we follow the evolution of a virtual subhalo and calculate its orbit. We assume that virtual subhaloes follow an isothermal profile truncated at $R_{\rm 200}$. We first compute the dynamical friction using the prescription derived by~\citet{binney08}: 
\begin{eqnarray}\nonumber
&& \frac {{\rm d}\vec{v}}{{\rm d}t}_{\rm dynf} = -\frac{GM_{\rm sat}(t)}{r^{2}} {\rm ln}\Lambda \left( \frac{V_{c}}{v} \right)^{2} \\ 
  && \left\{ {\rm erf} \left( \frac{v}{V_{c}}\right) - \frac{\sqrt{\pi}}{2} \left( \frac{v}{V_{c}} \right) {\rm exp} \left[ - \left( \frac{v}{V_{c}} \right) ^{2}\right] \right\} \vec{e_{v}},
\label{eqn:dynf}
\end{eqnarray}
where $M_{\rm sat}$ is the mass enclosed within $R_{\rm 200}$ for the satellite halo, $r$ is the distance from the centre of the main halo within which the satellite resides, ${\rm ln}\Lambda$ is the Coulomb logarithm with $\Lambda=1+M_{\rm halo}/M_{\rm sat}$ as formulated by~\citet{springel01}, $V_{\rm c}$ is the circular velocity of the main halo at $R_{\rm 200}$, and $v$ is the orbital velocity of the subhalo. When the subhalo reaches $0.1R_{\rm 200}$ under this prescription($t_{0.1R_{200}}$), the galaxy in the subhalo is considered to have merged into the central galaxy of the main halo. 

We also calculate a virtual subhalo's merger timescale, $t_{\rm merge}$, using the following fitting formula introduced by~\citet{jiang08}:
\begin{eqnarray}
t_{\rm merge} {\rm (Gyr)}= \frac{0.94\epsilon^{0.60}+  0.70}{\ln [1+(M_{\rm halo}/M_{\rm sat})]} \frac{M_{\rm halo}}{M_{\rm sat}} \frac{R_{\rm 200}}{V_{c}},
\label{eqn:jiang}
\end{eqnarray}
where $\epsilon$ is the orbital eccentricity of the satellite, $M_{\rm halo}$ and $M_{\rm sat}$ are the $M_{\rm 200}$ of the halo and subhalo, $R_{\rm 200}$ is the radius of the main halo, and $V_{\rm c}$ is the circular velocity of the main halo at $R_{\rm 200}$. We assume the galaxy in the virtual subhalo merges with the central galaxy at the smaller timescale between $t_{0.1R_{200}}$ and $t_{\rm merge}$.

Tidal stripping leads to mass loss from subhaloes in dense environments. We calculate the radius at which the gravitational force exerted by a virtual subhalo and its main halo are equivalent, the so-called sphere of influence, as follows~\citep{battin87}: 

\scriptsize
\begin{eqnarray}
 r_{\rm soi} \sim r  \left[ \left( \frac{M_{\rm sat}}{M_{\rm halo} (<r)} \right)^{-0.4}(1+3\cos ^2 \theta)^{0.1}+0.4\cos \theta \left( \frac{1+6\cos ^{2} \theta}{1+3\cos ^{2} \theta}\right) \right] ^{-1},
\end{eqnarray}
\normalsize
where $r$ is the distance between the centres of the virtual satellite and its main halo, $M_{sat}$ is the  $M_{\rm 200}$ mass of the virtual satellite halo, $M_{halo} (<r)$ is the total (dark matter+baryon) mass of the main halo within $r$, and $\theta$ is the angle between the line connecting a particle in the virtual satellite to the centre of the virtual satellite halo and the line connecting the centres of the virtual satellite and its main halo. We assume that dark matter outside this radius is tidally stripped on the dynamical timescale of the virtual subhalo, and the density profile of the virtual subhalo is instantly relaxed following the suggestion for a modification of the NFW profile by~\citet{hayashi03}.

Finally, when a satellite galaxy merges with the central galaxy, not all of its stellar mass is added to the central galaxy. Observations show that the stellar halo surrounding our own Galaxy is composed in part from the remains of destroyed satellites. Following observations for intra-cluster light in groups or clusters~\citep[e.g.][]{feldmeier02,gonzalez05,zibetti05} and theoretical studies of their origins~\citep{murante04,monaco06}, we scatter some stellar components from satellite galaxies over the main halo in our model. These components make up diffuse intracluster light in groups and clusters. The details of how we do this is not important for this study, so we refer interested readers to \citet{lee13}

\begin{figure*}
\centering 
\includegraphics[width=0.95\textwidth]{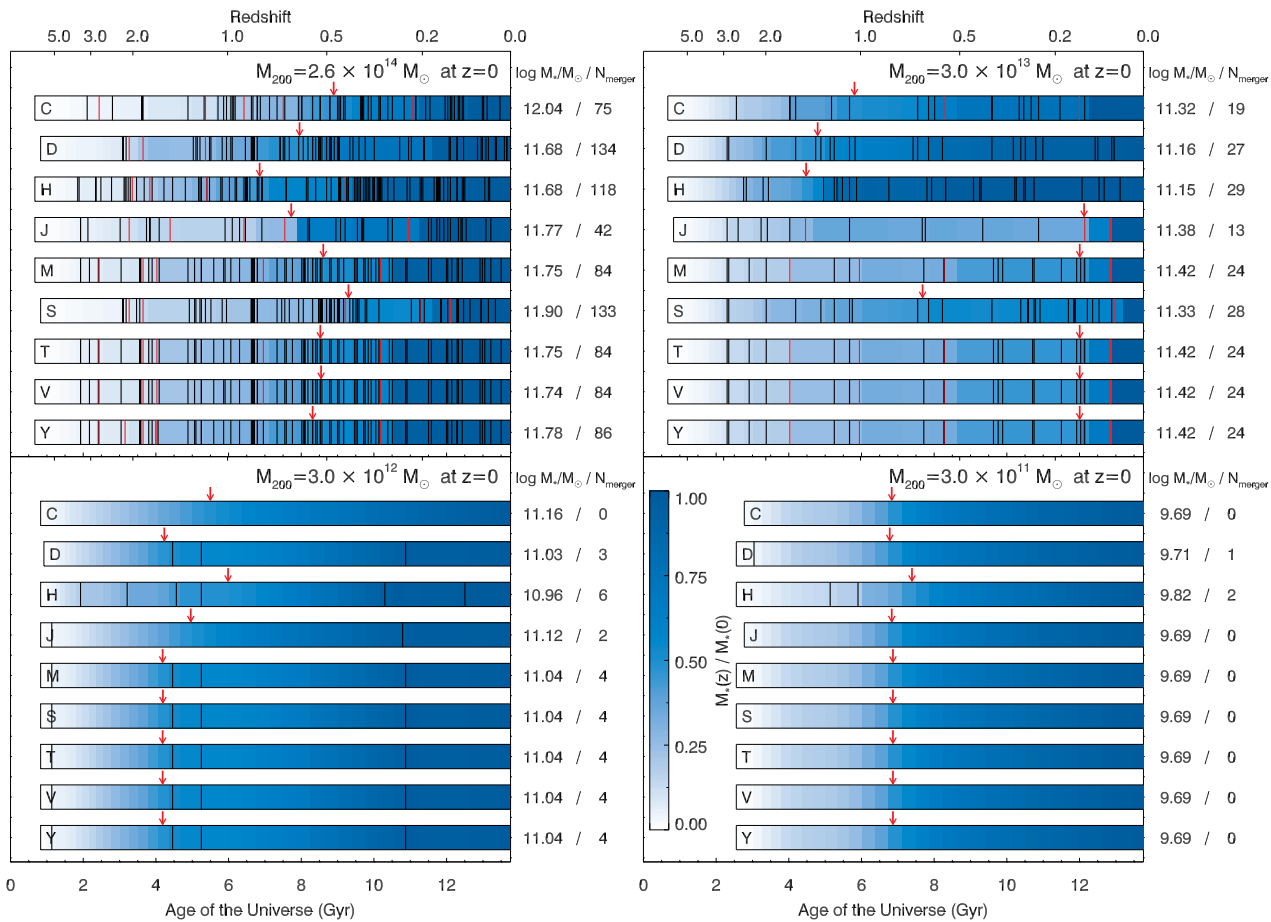}
\caption{Examples of the mass growth histories for central galaxies in four dark matter haloes. The first value on the right side of the panels shows the final stellar mass of the galaxies on a log scale at $z=0$. The second value presents the total number of mergers. The color gradient describes the linearly normalized mass according to the final stellar mass. The vertical lines indicate the epoch at which the galaxies are involved in mergers and the red lines show major mergers($\mu > 0.25$).  The red arrow indicates the epoch when the stellar mass of the galaxy reached half of the final stellar mass. The character at the left hand end of the bar represents the algorithm: `C'onsistent Trees, `D'-Trees, `H'BT, `J'Merge, `M'ergerTree,  `S'ubLink, `T'reeMaker, `V'ELOCIraptor, and `y'SAMtm}
\label{history}
\end{figure*}

\subsection{Lighting up the universe: Prescriptions for baryonic physics}
Here we briefly summarize the baryonic physics implemented in our SAM. Further details can be found in~\citet{lee13}. Our model calculates the cooling of hot gas based on the models proposed by~\citet{white91} and~\citet{sutherland93}. We assume that stars are formed in disks from a cold gas component, according to the simple law proposed by~\citet{kauffmann93}. We enable merger-induced starbursts in our model by adopting the prescription in~\citet{somerville08}, which is formulated using the hydrodynamic simulations performed by~\citet{cox08}. In addition, our model takes into account feedback processes. The prescription for supernova feedback comes from~\citet{somerville08}. Our SAM includes both quasar-mode and radio-mode AGN feedback, following the prescriptions proposed by~\citet{kauffmann00} and~\citet{croton06}, respectively. Our model also addresses some of the environmental issues that can affect the gas component of subhaloes; for instance, the hot gas component can be stripped by tidal forces~\citep[see][]{kimm11} and ram pressure~\citep{mccarthy08,font08}. 

\section{Galaxies from the halo merger trees}

Different halo growth histories give rise to different galaxy merger and gas accretion histories. Thus, the effect of varying the halo merger trees is evident when we examine the growth history of an individual galaxy. Figure~\ref{history} shows examples of central galaxy mass growth histories derived from nine different halo merger tree building algorithms working on a common halo catalogue. Even though the $z=0$ differences in stellar mass of these galaxies are small, the main haloes have different growth histories resulting in different evolutionary histories. We find that the galaxies grown in the trees extracted from similar algorithms naturally have similar properties. Thus, \mergertree, \treemaker, \velociraptor, and \ysamtm, which all have similar algorithms, result in approximately the same galaxy merger histories, as well as final stellar mass in all cases. On the other hand, \consistenttree, \dtree, \hbt, \jmerger, and \sublink\ show distinctive differences in the mass range where galaxy mergers are most frequent.

In this section, we investigate the impact of different halo merger tree building algorithms on global galaxy properties (as aggregates of individual galaxy properties), from the nine tree builders shown in Figure~\ref{algorithm}.

\subsection{Results from a common parameter set}
In order to cleanly evaluate the differences induced by the trees, we ran our SAM using a common parameter set. The parameter set was initially optimized for the \ysamtm\ tree that we developed for the semi-analytic model used in this study. Thus, the differences between the trees shown in this section do not indicate which tree is better or worse.

\subsubsection{Star formation history}

\begin{figure*}
\centering 
\includegraphics[width=0.95\textwidth]{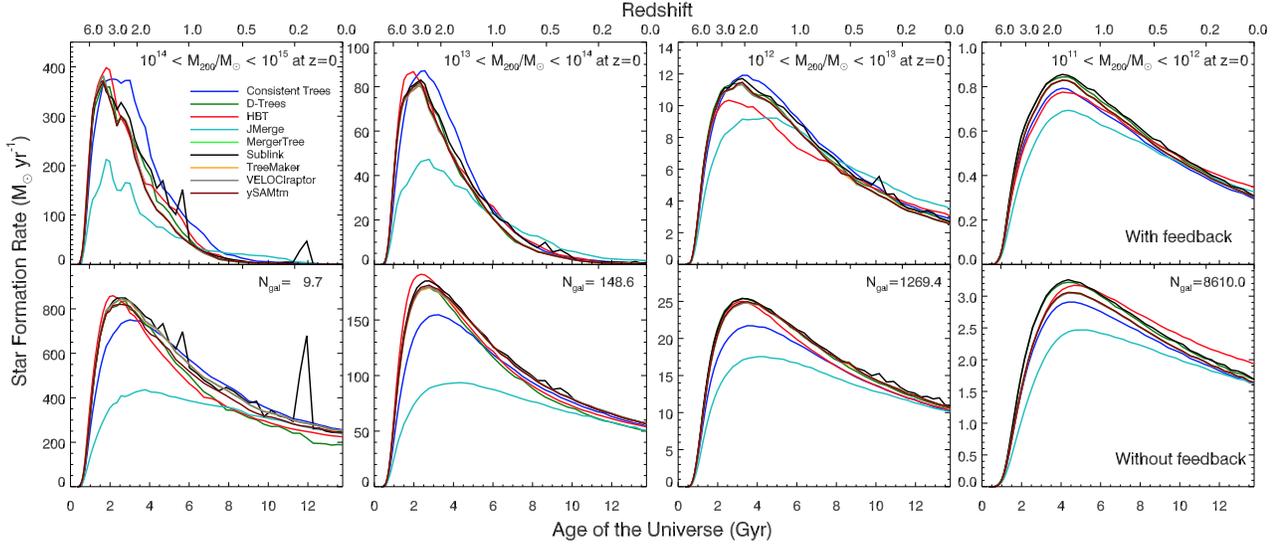}
\caption{ Star formation histories of main galaxies with respect to the $M_{\rm 200}$ of haloes. The upper and lower panels show models with and without feedback, respectively. The color coding, consistent throughout \citetalias{srisawat13} and this paper, represents the nine algorithms. ${\rm N}_{\rm gal}$ shows the mean number of main galaxies, averaged over different trees in each halo mass range. The upper panels have the same number of galaxies. }
\label{sfr2}
\end{figure*}

In semi-analytic approaches an initial hot gas reservoir, which is the main source of cold gas, is seeded using the cosmic baryonic fraction ($\Omega_{\rm b}/\Omega_{\rm m}$) and the halo mass. Baryons additionally flow into haloes via smooth accretion, which we assume is shock heated to the virial temperature of the main halo. Sometimes, galaxies can lose their hot gas if the halo harbouring them undergoes mass loss. Thus, if there are no feedback processes capable of disturbing the gas, the growth history of a halo strongly correlates with its star formation history. Figure~\ref{sfr2} shows the evolution of the mean star formation rates (SFRs) of galaxies living in main haloes within the given mass ranges at $z=0$ with and without feedback. In both cases, it is apparent that there are four distinct groups, classified according to the SFR histories. \consistenttree, \hbt, and \jmerger\ have distinct features from the other six trees. Interestingly, the trees with similar SFRs are also based on algorithms that are similar as shown in Figure~\ref{algorithm}. As seen in all the panels, the mean SFRs of \jmerger\ are always lower and peak later than the others. The spike on the SFRs of \sublink, shown in the bottom left panel, appears due to a starburst induced by a major merger. The starburst is conspicuous in the mean SFRs because there are only 10 galaxies in the mass bin and galaxies without feedback have more cold gas. Thus, one should not pay much attention to the spike and other wiggles in the most massive case (first column in Figure~\ref{sfr2}).

\begin{figure}
\centering 
\includegraphics[width=0.45\textwidth]{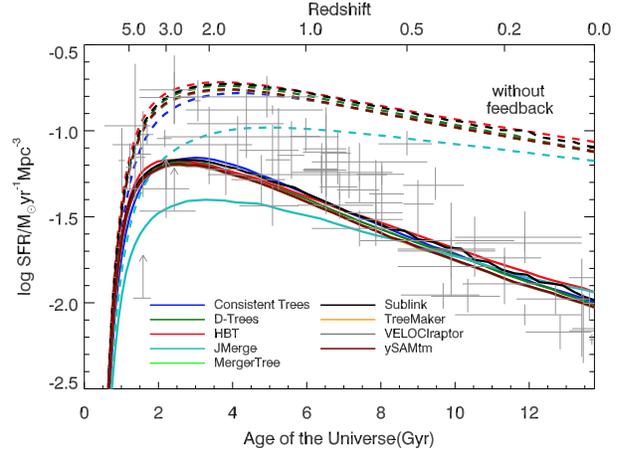}
\caption{Global star formation history. The solid and dashed-dotted lines show feedback models based on different halo merger trees as indicated by the key. The dashed lines represent models without feedback. The gray crosses indicate the empirical data compiled and modified by~\citet{panter07}}
\label{gsfr1}
\end{figure}

Across all of the tree building algorithms and mass ranges, feedback processes reduce the SFRs to less than half of the no feedback values (note the different y-axis scales). Supernova feedback takes effect in small haloes, while AGNs become important at larger halo masses. The amount by which the SFRs are suppressed increases as the halo masses are lowered, from a factor of two for high mass haloes to a factor of three for the smaller haloes. This is entirely in line with previous work and required in order to produce galaxies with realistic stellar masses.

On the other hand, the central galaxies in massive haloes ($M_{\rm 200} >10^{13}M_{\odot}$) experience a sharp decline in SFR over time due to AGN feedback which effectively shuts off the star formation at $z<0.5$. It has been suggested that super massive black holes (SMBHs) grow primarily via mergers~\citep{hernquist89, dimatteo05, springel05b, hopkins08, schawinski10}. This process activates a quasar feedback mode which eventually terminates star formation by violently blowing away cold gas due to high accretion rates onto the SMBH. Furthermore, SMBHs can persistently suppress star formation by radio mode feedback which is turned on at low gas accretion rates~\citep[e.g.][]{binney95, churazov02, fabian03, binney04, omma04, dubois10}. Because the accretion rate in the prescription for the radio mode AGN feedback in our model is proportional to the cube of the virial velocity~\citep{croton06}, central galaxies in more massive haloes are expected to show stronger radio mode feedback. However, SMBHs in small haloes do not accrete material at a high enough rate due to a low merger rate for AGN feedback to be the main channel of quenching star formation in this mass range.

Figure~\ref{gsfr1} presents the evolutionary histories of the global star formation rates (GSFRs) with and without feedback. There are trends here that are similar to those seen in Figure~\ref{sfr2}. The GSFRs of \jmerger\ are always lower than the others. The star formation histories with feedback processes follow observations well, except for that of \jmerger, which has the lowest peak amongst all of the trees. Although the peak SFRs and the epochs at which the trees reach the peaks differ from each other, the final SFRs at $z=0$ are almost the same between the trees. This is mainly because the differences in halo growth history between the algorithms is prominent at early epochs when haloes are small and their mergers are frequent. Thus, the uncertainty of progenitor-descendant relation between the trees is bigger in earlier epochs. On the other hand, the recent growth history of haloes does not differ much between the trees.

\begin{figure}
\centering 
\includegraphics[width=0.45\textwidth]{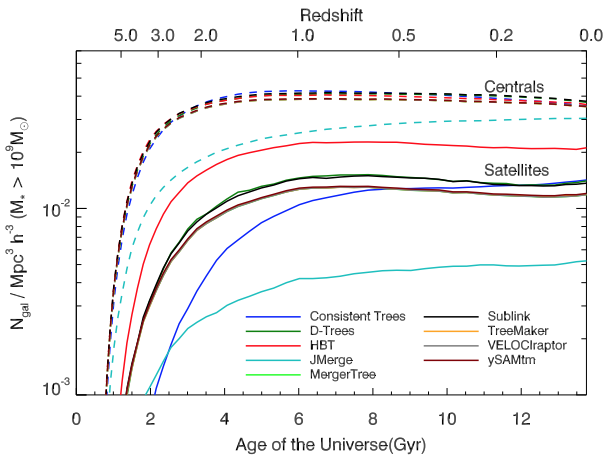}
\caption{The number density evolution of galaxies more massive than $10^{9}M_{\odot}$ in a comoving volume. The dashed lines represent the number density of the central galaxies. The solid lines display the evolutionary history of the satellite number density.}
\label{ngal_evo}
\end{figure}

\subsubsection{The number of galaxies}

As described in \S3.2, although all of the tree building algorithms used a common halo catalogue as a starting point, the number of trees and branches produced that are suitable for our semi-analytic model can vary. For instance, if there are missing links along a main branch the earlier progenitor haloes prior to the break were discarded by our semi-analytic model. Thus, if an algorithm has additional processes to minimise the segmentation of merger trees, it can result in a larger number of galaxies at an epoch than other algorithms do. Figure~\ref{ngal_evo} shows the number density evolution of galaxies more massive than $10^{9}M_{\odot}$ in a comoving volume for all of the tree builders. As expected the number density of the satellite galaxies varies, while that of the main galaxies is similar, even at high redshifts. 

Additionally, our semi-analytic model calculates the orbits of the subhaloes that disappear from a halo catalogue before reaching the central region of the main halo. Therefore, the number of satellite galaxies in the models is not always the same as the number of subhaloes that was initially provided by the halo catalogue. Among all of the trees, \hbt\ has the largest number of satellite galaxies at $z=0$, followed by \consistenttree, \dtree\ and \sublink. We are barely able to distinguish between those of \mergertree, \treemaker, \velociraptor, and \ysamtm\ which are all very similar tree building algorithms. \jmerger\ has the smallest number of satellites. As the star formation history implies (Figure~\ref{sfr2}), \jmerger\ tends to have the shortest halo growth histories among all of the tree building routines (Figure~\ref{ngal_evo}). It barely reaches the number density of central galaxies of other algorithms by $z=0$, and the situation is even more striking for satellite galaxies. While \consistenttree\ does not have a distinguishable history in terms of the number density of main galaxies the story is not the same when it comes to satellites. Even though it has the lowest number density at $z>3$, it surpasses \jmerger\ with a rapid increase and ends up as second at $z=0$. This is because \consistenttree\ prunes haloes that show large mass fluctuation, considered to be the outcome of misidentification by halo finders. Subhaloes at high redshifts are more likely to undergo mass fluctuation due to their low mass, and thus many of them are removed by the algorithm. However, at the mass resolution of this model, there are 5 to 10 times more central than satellite galaxies at all redshifts and thus the large difference in the number of satellites does not cause a noticeable difference in the galaxy mass function as will be discussed later.

\begin{figure}
\centering 
\includegraphics[width=0.45\textwidth]{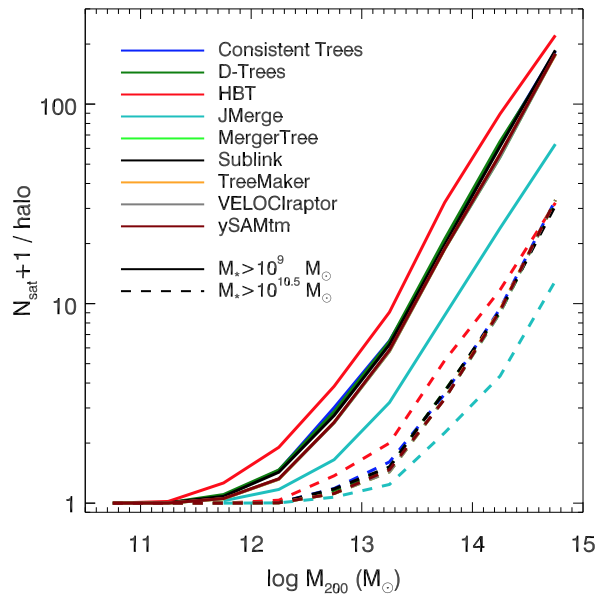}
\caption{ The mean number of satellite galaxies that are more massive than $10^{9}M_{\odot}$(solid) and $10^{10.5}M_{\odot}$(dashed) within a main halo with the indicated halo mass at $z=0$.}
\label{ngal}
\end{figure}

\begin{figure*}
\centering 
\includegraphics[width=0.85\textwidth]{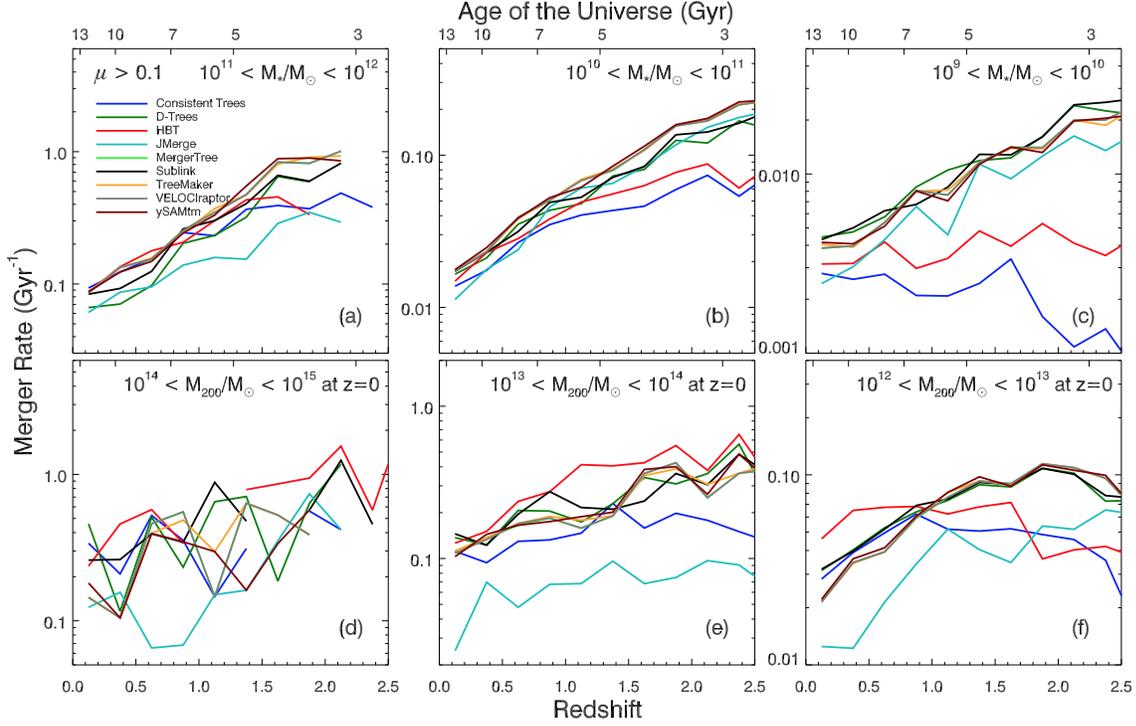}
\caption{ The mean merger rates (mergers per galaxy per Gyr) of galaxies according to galaxy stellar mass at each epoch (upper panels) and the merger rate evolution of central galaxies finally hosted by haloes in a given mass range at $z=0$ (lower panels). The mass ratio cut adopted, $\mu = M_{2}/M_{1}$, is 0.1.}
\label{merger1}
\end{figure*}

Figure~\ref{ngal} presents the mean number of satellite galaxies with stellar masses $>10^{9}M_{\odot}$ and $>10^{10.5}M_{\odot}$ per halo plotted relative to the main halo's $M_{\rm 200}$ at $z=0$. \hbt\ has more satellites than the other tree building methods by virtue of its algorithm, which was developed specifically to track subhaloes. On the other hand, due to its algorithm, \jmerger\ has the smallest number of satellites. Therefore, \jmerger\ loses many of its subhaloes when it builds merger trees. The other seven trees have almost the same number of satellite galaxies at all main halo masses.

\subsubsection{Galaxy merger history}

In this section, we describe our analysis of galaxy merger history and its contribution to the final galaxy mass. Figure~\ref{merger1} shows the evolution of the mean merger rates of galaxies according to stellar mass for all epochs (top panels) and the history of the mean merger rates of central galaxies that end up in a given halo mass at $z=0$ (bottom panels). Because galaxy mergers is one of the channels that affects galaxy stellar mass, different merger rates lead to different compositions of the stars in terms of their origins. There are some missing links in the most massive cases as seen in the bottom panels, and this is due to the lack of massive haloes. 

\begin{figure}
\centering 
\includegraphics[width=0.45\textwidth]{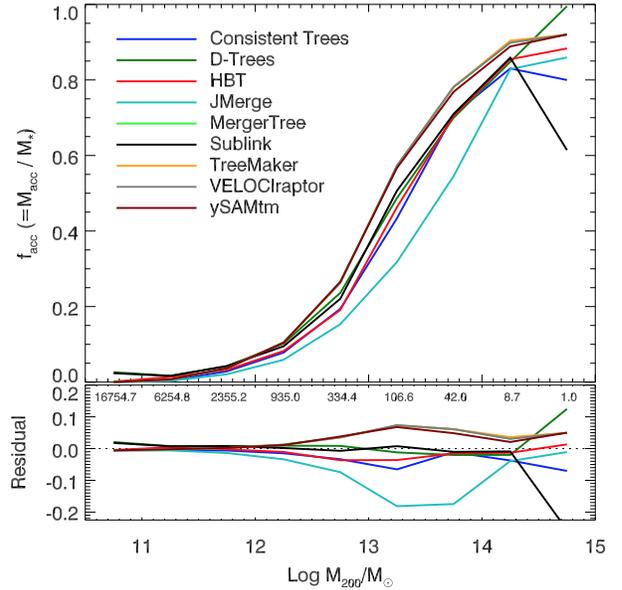}
\caption{ The mean contribution of merger accretion to final stellar mass according to the $M_{\rm 200}$ of main haloes at $z=0$. The residuals show the differences between the mean value of all the models and the fractions of the individual algorithms. The numbers in the bottom panel present the mean number of haloes at each mass bin, with intervals of 0.5 dex from 10.5.}
\label{accr}
\end{figure}
 
\consistenttree, \hbt\, and \jmerger\ show notable differences for different reasons. First, \consistenttree\ trims haloes showing large mass fluctuations, as described in 4.1.2, which occur more frequently in dense environments. Hence, subhaloes in dense environments are often removed by this algorithm, which results in lower merger rates than others in some conditions (e.g., panels a, b, c, e, f). \hbt\ on the other hand shows lower merger rates (panels b, c, f) because subhaloes in \hbt\ have longer lifetime than those in the other trees.  \citetalias{srisawat13} demonstrate that haloes in \hbt\ have the fewest direct progenitors among all the algorithms. If a tree building algorithm fails to trace a branch of a subhalo, then the subhalo would be regarded as being merged into its main halo as one of direct progenitors. Accordingly, better algorithms in terms of tracking subhalo branches would have fewer direct progenitors on average. With its algorithm allowing itself to rigorously build the branches of subhaloes, \hbt\ has the highest satellite number density across cosmic time, as shown in Figure~\ref{ngal_evo}, inevitably accompanying lower merger rates. \jmerger\ in the upper panels of Figure~\ref{merger1} does not appear significantly different from other trees. It is a result of two facts about \jmerger: there are fewer mergers but there are also fewer galaxies in a unit volume. In bottom panels (in particular in panels e and f), \jmerger\ shows remarkably lower merger rates due to the fact that main haloes of \jmerger\ have fewer branches than those of the other trees. In the most massive cases (panel d), however, all the algorithms are similar to each other because subhaloes haboring satellite galaxies with $\mu > 0.1$ in massive haloes are also massive enough to be well tracked, regardless of the algorithms.

The different galaxy merger rates naturally result in varying stellar accretion rates. Figure~\ref{accr} shows the fraction of stellar mass accreted by central galaxies from mergers according to halo mass at $z=0$. Central galaxies in massive haloes $(\geq10^{14}M_{\odot})$ accrete $\geq80\%$ of their stars via mergers. This fraction rapidly falls with decreasing halo masses. With its low merger rates in the bottom panels of Figure~\ref{merger1}, \jmerger\ has the smallest fraction. Note that the most massive bin ($M_{\rm 200}>10^{14.5} M_{\odot}$) contains only one halo and thus is vulnerable to stochastic effects. For example in the case of \dtree, the central galaxy of the sole most massive bin has less than $10^{9}M_{\odot}$ of stars that are born in-situ. \sublink\ shows a sudden break at the most massive range. It happens because the central galaxy undergoes a big starburst due to a gas rich merger; it is just a stochastic effect as well. In short, the merger accretion rates, $f_{\rm acc}$, tightly correlates with halo mass mainly because the halo merger rate increases with halo mass. 

\begin{figure}
\centering 
\includegraphics[width=0.45\textwidth]{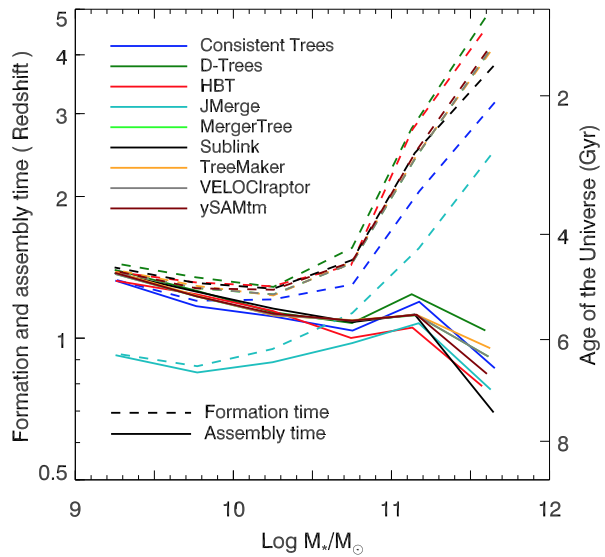}
\caption{ The mean formation and assembly time of main galaxies (the times by which 50\% of the stars have formed/assembled into the progenitor of the final object) according to their final stellar mass. The dashed lines indicate the formation time of the stellar components initially born in all progenitors. The solid lines indicate the epochs at which the stellar components are finally assembled within the main branches.}
\label{half}
\end{figure}

\subsubsection{Stellar mass assembly time}

In previous sections, we discussed the mechanisms that increase galaxy stellar mass. The star formation histories shown in Figure~\ref{sfr2} reveal that, for our SAM at least, galaxies in more massive haloes are born earlier and grow faster than those in less massive haloes. Figures~\ref{merger1} and~\ref{accr} clearly demonstrate that galaxies in more massive haloes are involved in more mergers, even at low redshifts, and thus acquire stellar mass predominantly from mergers. 

In this study, the `formation time' is the time by which half of the stars which end up in the galaxy at $z=0$ have formed. The `assembly time' of a galaxy indicates the epoch at which the stellar mass assembled into the central object reaches half the final stellar mass. If a galaxy grows only through in-situ star formation, its formation time and assembly time should be the same. On the other hand, if galaxy mergers continuously supply additional stellar components to a galaxy, the assembly and formation time of the stellar components can be different. In practice, as galaxies do not form monolithically, a galaxy's formation always occurs before its assembly time.

Figure~\ref{half} shows the formation and assembly time of galaxies according to their final stellar mass. As the concept of downsizing stipulates~\citep{cowie96,glazebrook04,cimatti04}, stars in more massive galaxies tend to be formed earlier than those in less massive ones (dashed lines). This trend is shown in Figure~\ref{sfr2}. When the stellar mass of the main branches (direct progenitors) increases, however, it takes longer to reach half of its final stellar mass. More massive galaxies are more likely to obtain their mass via accretion through mergers, and the merger accretion rates decay more slowly than the SFRs~\citep{oser10,cattaneo11,lackner12,lee13}. Thus, with increasing mass, the time interval between the formation and assembly time increases. As seen in Figures~\ref{sfr2} and~\ref{gsfr1}, the galaxies in \jmerger\ are formed and assembled on average later than those from the other tree building algorithms. This is entirely in line with our assertion that \jmerger\ artificially truncates many trees when it fails to link two halos between successive snapshots.

\subsubsection{Galaxy mass at $z=0$}

\begin{figure}
\centering 
\includegraphics[width=0.45\textwidth]{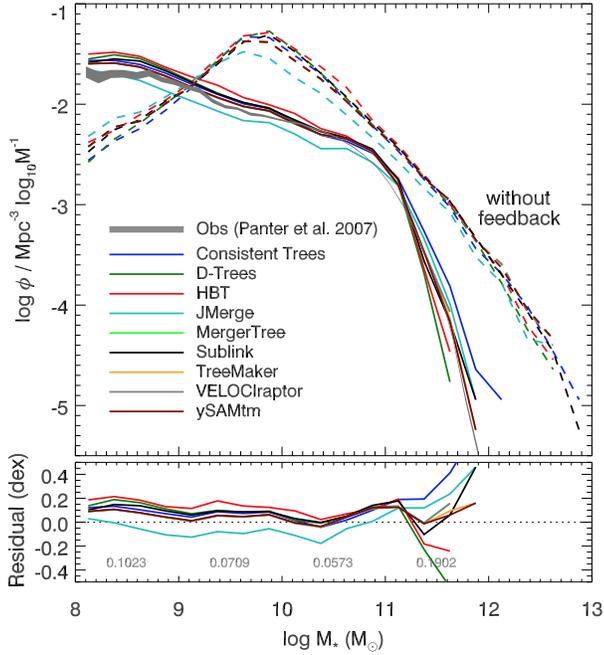}
\caption{ Galaxy stellar mass functions and the residuals between the models and the empirical data . The solid and dashed-dotted lines represent the galaxy stellar mass functions from models with various halo merger trees and the gray shading represents the empirical stellar mass function derived by~\citet{panter07}. The dashed lines show the galaxy mass functions of models without feedback. The bottom panel shows the residuals of the model mass functions with respect to the empirical mass function. The numbers in the panel display the mean absolute residuals at each mass bin.}
\label{mf1}
\end{figure}

The galaxy mass function, or luminosity function, is one of the important global galaxy properties. Although there can be degeneracy, the galaxy mass function reflects important physical processes involved in galaxy formation and evolution. Figure~\ref{mf1} shows the galaxy stellar mass functions from empirical data~\citep{panter07} and models, and the residuals between them. We also present a case without feedback using dashed lines. The residuals are not huge in the low mass region($M_{\rm 200}<10^{11}M_{\odot}$); however, they severely diverge from each other at the high mass end. Most of the low mass galaxies reside in small haloes, either alone or with a few companions. The channel through which they grow is quiescent star formation. Therefore, their growth histories may be far less complicated than those of galaxies in large haloes. Massive galaxies in big haloes, however, are in the forefront of all physical processes governing galaxy formation and evolution. We should note that the small number of massive haloes in our volume causes a stochastic effect, so the variance seen at the massive end should be treated with caution.

\begin{figure}
\centering 
\includegraphics[width=0.45\textwidth]{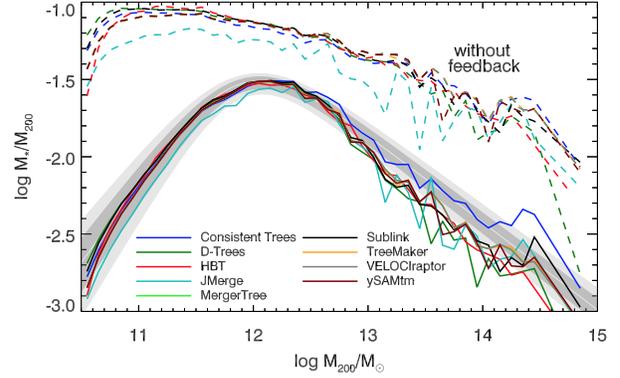}
\caption{The stellar to halo mass ratio with $M_{\rm 200}$ at $z=0$. The white solid line shows the mean ratio derived by~\citet{moster10}, and the dim and bright gray shades indicate $1\sigma$ and $2\sigma$ ranges, respectively. The solid and dashed-dotted lines come from models with feedback processes and the dashed lines indicate the cases with no feedback.}
\label{virm_star1}
\end{figure}

The differences between the models are not as pronounced in the no feedback case. However, both cases, with and without feedback, reveal a unique characteristic of the \jmerger\ algorithm: a short formation time. \jmerger\ has fewer galaxies between $10^{8}<M_{*}/M_{\odot}< 10^{11} $ for the feedback case than the other algorithms as seen in Figure~\ref{ngal_evo}. 

Feedback alters the simple proportional relationship between halo mass and stellar mass. Figure~\ref{virm_star1} presents the stellar to halo mass ratio of the empirical data and our models with and without feedback at $z=0$. All the trees with feedback produce galaxies that follow the empirical data of~\citet{moster10}, although the deviation and fluctuations increase with increasing mass. The stellar to halo mass ratios with and without feedback demonstrate how feedback processes work in our models. The gradually decreasing ratio with the increasing $M_{\rm 200}$ in the no-feedback case can be attributed to two factors. Firstly, the cooling efficiency of hot gas decreases with increasing halo mass. In addition, stars in massive haloes belong not just to the central galaxy, but to its satellites and the diffuse stellar halo. The case without feedback more directly demonstrates the effect of the tree building algorithms. As previously stated, \jmerger\ has the lowest stellar mass in all halo mass ranges, due to its short formation and assembly time. However, all the other tree building algorithms behave similarly. 

\begin{figure}
\centering 
\includegraphics[width=0.45\textwidth]{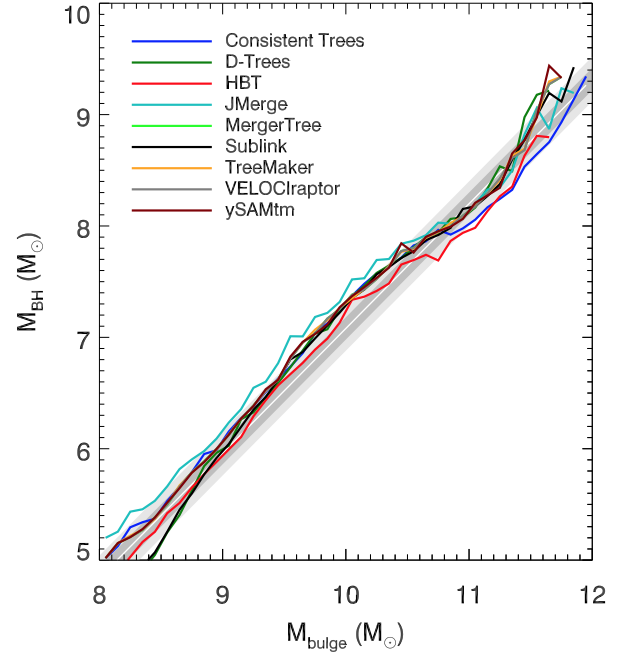}
\caption{Central supermassive black hole-to-bulge mass relation. The white solid line is the empirical relation derived by~\citet{haring04} and the dim and bright grey shading shows the 1 and $2\sigma$ confidence regions, respectively.  }
\label{mbh}
\end{figure}

\subsubsection{The $M_{\rm BH}$-$M_{\rm bulge}$ relation}

The $M_{\rm BH}$-$M_{\rm bulge}$ relation can be a simple calibration point. Figure~\ref{mbh} displays the $M_{\rm BH}$-$M_{\rm bulge}$ relations for our models and empirical data. The relations of the model galaxies are similar to the empirical data derived by~\citet{haring04} and also to each other. It is impossible to recognize distinct differences between the algorithms in essence due to the fact that the prescriptions used to feed SMBHs in our models allow for black hole growth mainly when galaxy mergers take place. If a tree building algorithm gives rise to more galaxy mergers, the galaxies governed by the algorithm have an increased chance of fueling their central galaxies, which would increase bulge stellar mass at the same time. Although the radio-mode AGN feedback also contributes to the growth of central black holes, the accretion rate of the radio-mode is far lower than that of the quasar-mode, which is turned on by mergers with enough cold gas. Thus, even the use of different merger tree algorithms simply moves model galaxies along a pre-defined locus in the $M_{\rm BH}$-$M_{\rm bulge}$ diagram.

\subsection{Results from parameter sets calibrated for each tree}

In the previous section, we detailed the effect of various tree building algorithms on galaxy properties using a semi-analytic model. In order to focus on the impact of halo merger trees, we adopted a common parameter set for all of the trees, initially derived using the \ysamtm\ tree builder. 

Here we find SAM parameter sets optimized for a given tree and examine the resulting differences not only in the galaxy distribution but the underlying physics characterised by the parameters of our SAM. The impact of merger tree appears larger on the most massive galaxies (Figure~\ref{mf1}). Thus, we search for the optimal parameter set for each model as follows. First, we find the parameter space that matches the overall shape of the lower-mass end ($M_{*}<10^{11}M_{\odot}$) of the mass function at $z=0$. Then, we choose a parameter set which minimizes the absolute residuals in the massive end ($M_{*}>10^{11}M_{\odot}$) of the mass function as a calibration.

\subsubsection{Galaxy mass at $z=0$}

\begin{figure}
\centering 
\includegraphics[width=0.45\textwidth]{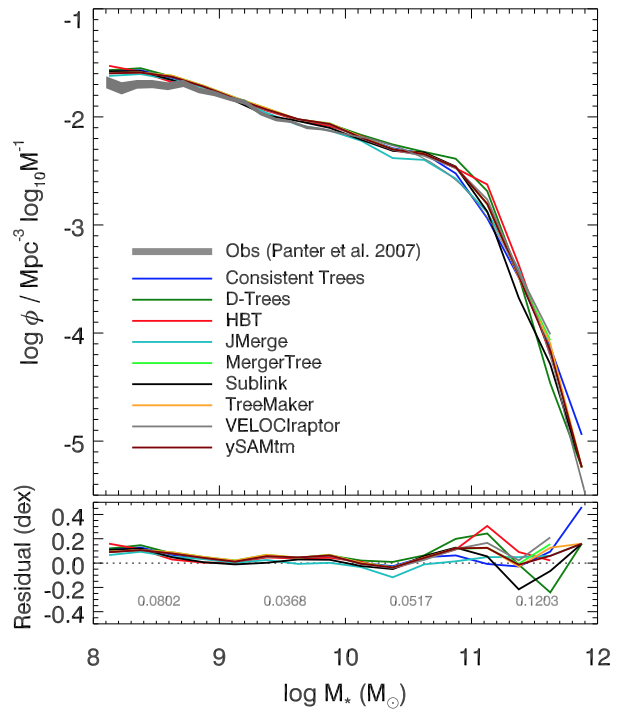}
\caption{Galaxy stellar mass functions and the residuals between the models individually calibrated for each tree algorithm and empirical data.}
\label{mf2}
\end{figure}

\begin{figure}
\centering 
\includegraphics[width=0.45\textwidth]{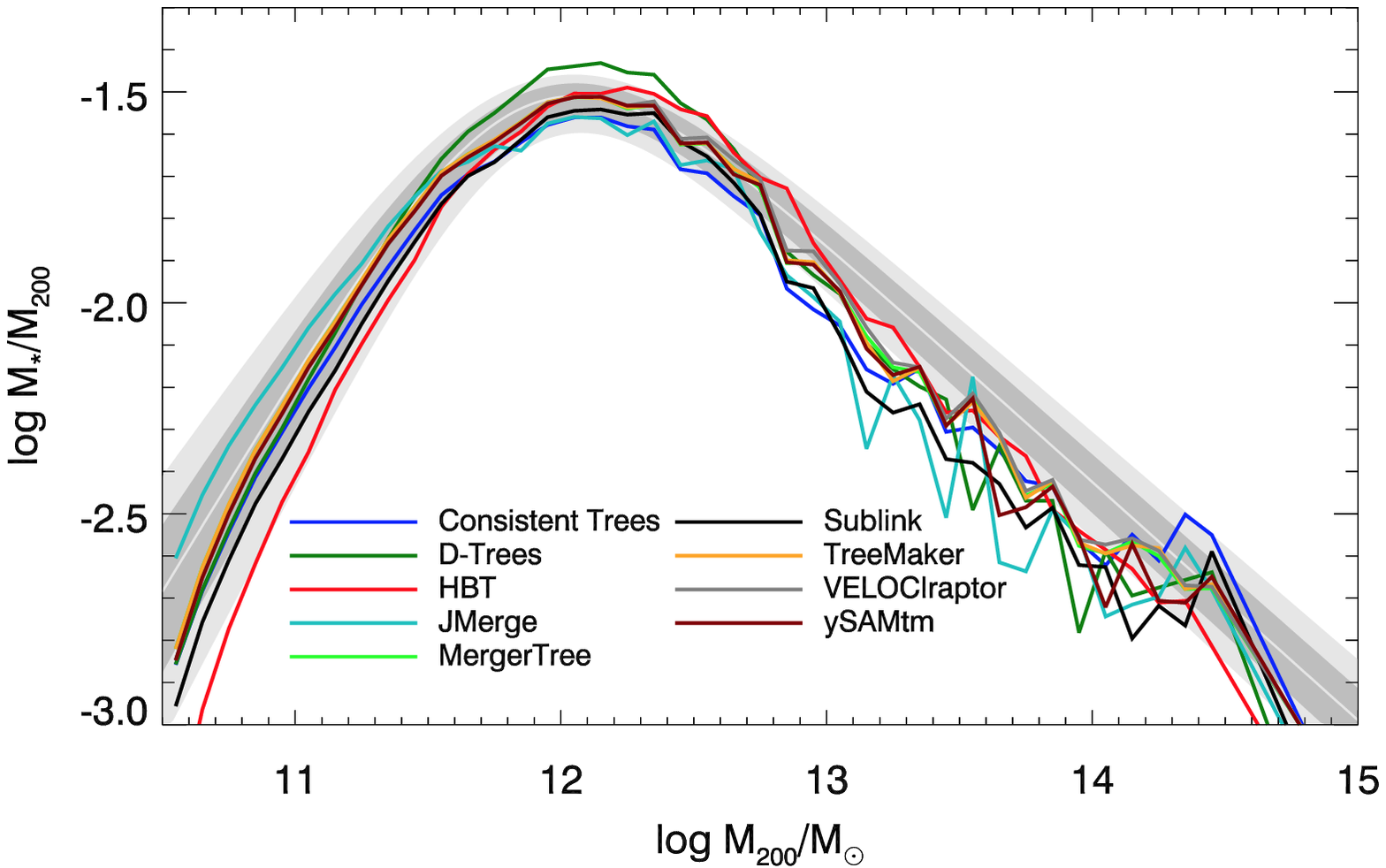}
\caption{Same as Figure~\ref{virm_star1}, but with individually calibrated (see text) models with feedback.}
\label{virm_star2}
\end{figure}

\begin{figure}
\centering 
\includegraphics[width=0.45\textwidth]{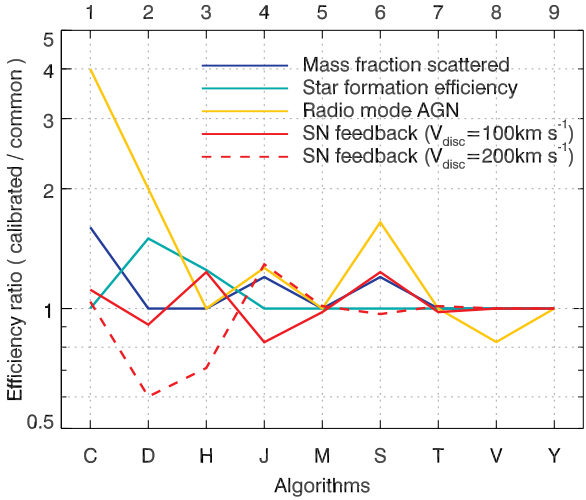}
\caption{ The efficiency ratio of physical processes between models calibrated for each tree algorithm and based on a common parameter set. The letters on the x-axis represent the algorithms: `C'onsistent Trees, `D'-Trees, `H'BT, `J'Merge, `M'ergerTree, `S'ubLink, `T'reeMaker, `V'ELOCIraptor, and `y'SAMtm }
\label{param}
\end{figure}

We calibrated the models to each tree by modifying the strength of four physical processes: star formation efficiency, the stellar mass fraction scattered due to mergers, radio mode AGN feedback, and supernova feedback efficiency.

We calculate star formation rates within cold gas discs using a simple formula proposed by~\citet{kauffmann93}:
\begin{eqnarray}
 \dot{m}_{*}=\alpha \frac{m_{\rm cold}}{t_{\rm dyn,gal}},
\label{quiescent}
\end{eqnarray}
where $\alpha$ is the star formation efficiency, $m_{\rm cold}$ is the amount of cold gas, and $t_{\rm dyn,gal}$ is the dynamical timescale of a cold gas disc. We adopt $\alpha=0.02$ as a default, but we can change this to adjust the overall star formation rate. 

It has been suggested that some of the stellar component of satellite galaxies are scattered by dynamical disturbances during mergers, and that these components finally belong to a diffuse stellar component in haloes (the intra-cluster light) rather than being simply added to the stellar mass of the central galaxy once the central and satellite galaxy have merged. This stellar mass, lost from the satellite galaxy, is described as follows:

\begin{figure*}
\centering 
\includegraphics[width=0.9\textwidth]{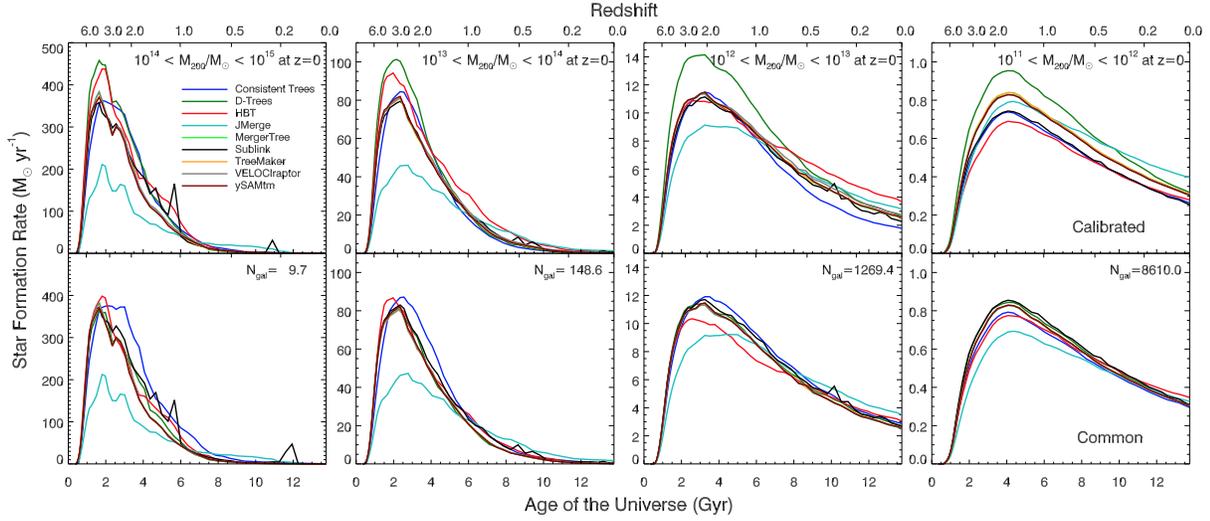}
\caption{Similar to Figure~\ref{sfr2}, however the top panels show the models with parameters optimized for each tree algorithm, whereas the bottom panels show the models with a common parameter set. }
\label{sfr3}
\end{figure*}

\begin{eqnarray}
M_{\rm scatter}=f_{\rm scatter}M_{\rm sat},
\label{scatter}
\end{eqnarray}
where $f_{\rm scatter}$ is the fraction of stars scattered by mergers and $M_{\rm sat}$ is the stellar mass of a satellite galaxy. Therefore, a higher value of $f_{\rm scatter}$ leads to more intra-cluster light and a lower final central galaxy mass.

The radio-mode AGN feedback has been adopted in many semi-analytic models to regulate the massive end of the galaxy mass function. We follow a prescription formulated by~\citet{croton06} to calculate the accretion rates in the radio-mode:
\begin{eqnarray}
 \dot{ m}_{\rm BH,R}=\kappa_{\rm AGN} \left( \frac{m_{\rm BH}}{10^{8}M_{\odot}} \right) \left( \frac{f_{\rm hot}}{0.1} \right) \left( \frac {V_{\rm vir}}{\rm 150km\, s^{-1}} \right)^{3},
\label{eqn:radio}
\end{eqnarray}
where $\kappa_{\rm AGN}$ is a free parameter with units of $M_{\odot}{\rm yr}^{-1}$, $m_{\rm BH}$ is the black hole mass, $f_{\rm  hot}$ is the mass fraction of hot gas in haloes, and $V_{\rm vir}$ is the virial velocity of haloes. Part of the mass accreted by the radio-mode turns into energy by $L_{\rm BH}=\eta \dot{m}_{\rm BH,R}c^{2}$, where $\eta$ is the conversion efficiency of rest mass to radiation, which is set to be 0.1 in general, and $c$ is the speed of light. We enhance or suppress the radio-mode AGN by tuning $\kappa_{\rm AGN}$.

Supernova feedback is effective on regulating the growth of small galaxies. We implement supernova feedback by adopting the prescription in~\citet{somerville08}:
\begin{eqnarray}
 \dot{m}_{\rm rh}=\epsilon^{\rm SN}_{\rm 0} \left( \frac{\rm 150km\,s^{-1}}{V_{\rm disc}} \right)^{\alpha_{\rm rh}} \dot{m}_{*},
\label{snf}
\end{eqnarray}
where $\dot{m}_{\rm rh}$ is the reheating rate of cold gas, $\epsilon^{\rm SN}_{\rm 0}$ and $\alpha_{\rm rh}$ are free parameters, $V_{\rm disc}$ is the rotational velocity of a disc, and $\dot{m}_{*}$ is the star formation rate. In this study, we use $V_{\rm 200}$ as a proxy of $V_{\rm disc}$, and modify the velocity criterion of Eq.~\ref{snf}.

Figures~\ref{mf2} \& \ref{virm_star2} show the stellar mass functions of the model galaxies and the stellar mass to halo mass, respectively, with parameter sets calibrated for each tree algorithm. The model mass functions in Figure~\ref{mf1} shows very small residuals in the low mass regions ($M_{\rm 200}<10^{10.5}M_{\odot}$). Therefore, we focused on calibrating them to reproduce the high mass end. Comparing to Figure~\ref{mf1}, the mean absolute residual in each mass bin slightly decreases with the calibrations. Both the mass function and stellar mass to halo mass of the trees are now in better agreement with the empirical data, though the stellar to halo mass ratio still remains slightly lower than the mean empirical values from~\citet{moster10} at the massive end.

Figure~\ref{param} shows the results of calibrating the SAM prescriptions for each tree algorithm individually. Here we show the efficiency ratios of the four processes between the models based on calibrated and common parameter sets: for instance, $ \dot{m}_{\rm rh}({\rm calibrated})/ \dot{m}_{\rm rh}({\rm common})$. Since the prescription for the supernova feedback, Eq.~\ref{snf}, depends on the rotational velocity of discs, we plot the efficiency ratios for two cases, $V_{\rm disc}=100{\rm km\,s}^{-1}$ and $200 {\rm km\,s}^{-1}$. As Figure~\ref{param} demonstrates, the massive end of \consistenttree\ is suppressed with stronger radio mode AGN feedback and a higher fraction of stellar mass in satellites scattered during mergers. There are galaxies that are more massive than $10^{12} M_{\odot}$ in Figure~\ref{mf1}; however, for the new calibrations, the most massive one produced by \consistenttree\ is $\approx7\times10^{11} M_{\odot}$. 
\dtree, which had few massive galaxies in the common case, is now in better agreement due to a higher star formation efficiency and weaker supernova feedback. The efficiency of radio mode AGN feedback is enhanced to balance the increase in the galaxy stellar mass.

As the massive objects are suppressed, the stellar to halo mass ratio of \consistenttree\ is also reduced. That of \hbt\ and \sublink\ was lowered in the low mass region, as their mass function was tuned with stronger supernova feedback in low $V_{\rm disc}$, correlating with low $M_{\rm 200}$ in general. On the other hand, \jmerger\ is enhanced in the low mass region by weaker supernova feedback efficiency in low $V_{\rm disc}$. The higher star formation efficiency and overall decrease of supernova feedback boost \dtree\ around $10^{12}M_{\odot}$.

\subsubsection{Mass growth history}

Figure~\ref{sfr3} shows SFRs from the best parameter sets compared to SFRs with a common parameter set, shown in the second row. Figure~\ref{mf1} demonstrated that \consistenttree\ has more massive galaxies than is seen in the empirical data. The new parameter set with stronger feedback and lower star formation efficiency suppresses star formation. On the other hand, \dtree, which had a deficiency of massive galaxies, has much higher SFRs for haloes of all masses. \hbt\ had slightly fewer galaxies in $M_{*}>10^{11}M_{\odot}$ and more in $M_{*}<10^{11}M_{\odot}$. The parameters for \hbt\ were modified in order to diminish SFRs in the low $M_{\rm 200}$ range and to increase them in the most massive range. \jmerger, which has fewer small galaxies than any of the other algorithms, undergoes an enhancement of SFRs in the low mass region. \sublink\ is able to decrease its low mass galaxy number by suppressing the SFR of the galaxies in the low $M_{\rm 200}$ range. The other trees have parameters similar to our fiducial set so they do not show any notable differences with the new parameters.

\begin{figure}
\centering 
\includegraphics[width=0.45\textwidth]{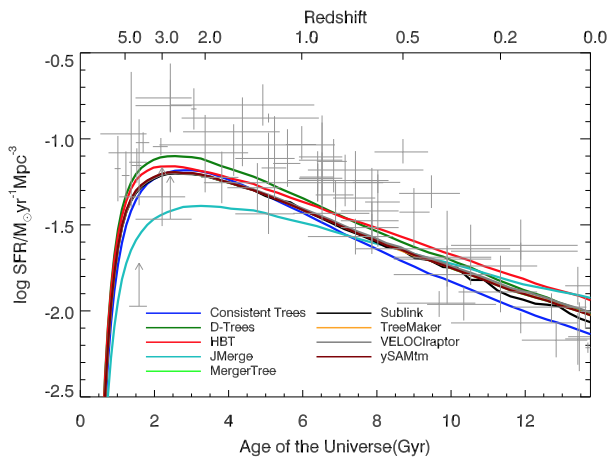}
\caption{Same as Figure~\ref{gsfr1}, but with individually calibrated (see text) models with feedback.}
\label{gsfr2}
\end{figure}

The new global star formation rates are shown in Figure~\ref{gsfr2}. Compared to the common parameter case (Figure~\ref{gsfr1}), \dtree\ and \jmerger\ have increased noticeably at high- and low-redshifts, respectively. 
Overall, the new models individually calibrated to match the galaxy mass function better show a larger variety of star formation history compared to the common-parameter case (Figure~\ref{gsfr1}).

\begin{figure}
\centering 
\includegraphics[width=0.45\textwidth]{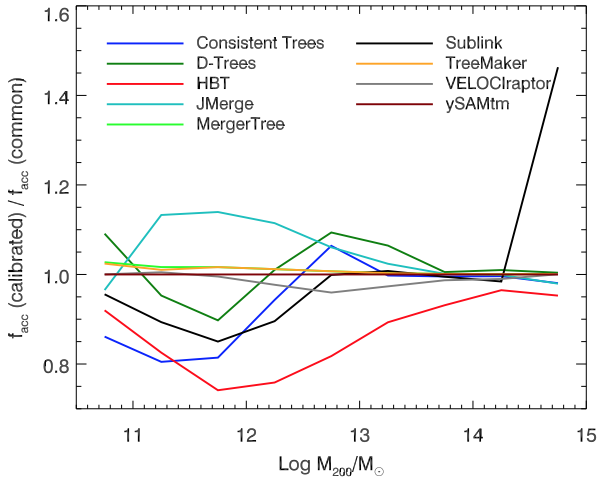}
\caption{ The ratio between the accretion fractions from models based on the common and new parameter sets. }
\label{accr2}
\end{figure}

The new models produce different individual galaxy growth histories. Therefore, we expected that the amount of stars accreted by the mergers would also change. Figure~\ref{accr2} shows the ratio between the accretion fractions in Figure~\ref{accr} and those in the new models. The ratio of \consistenttree\ is almost equal to unity for $M_{\rm 200}>10^{12.5}M_{\odot}$, then drops to below unity for smaller mass haloes. \dtree\ and \sublink\ show a similar trend. However, the drop in the accretion rate at small masses may not be significant as the accretion stellar mass fraction amounts to less than a few percent for low mass haloes. \sublink\ shows a dramatic sudden jump at $10^{14.5}M_{\odot}$. First, we note that it is derived from just one galaxy in the bin and thus subject to stochastic effect. In this particular case, the galaxy had significantly different {\em in-situ} star formation histories between common-parameter and individually-calibrated cases due to subtle changes in the mass ratios of merging galaxies which affect the ``wetness" of mergers. It is difficult to judge at the moment whether such a sensitivity is realistic or a computational artifact. The fraction for \hbt\ is lower than unity across the whole mass range. In Figure~\ref{sfr3}, the SFRs of \hbt\ are lower in low mass regions and slightly higher at the massive end when compared to the original parameter set due to enhanced supernova feedback and higher star formation efficiency. Thus, smaller galaxies are more strongly suppressed than before, resulting in lower accretion rates. Calibrating \jmerger\ required higher SFRs in the low mass regions; thus, its accretion rates have increased.

\section{Summary and Discussion}

\citetalias{srisawat13} discussed the differences between the halo merger trees built by various algorithms. Here we investigated the impact on the results of a semi-analytic model due to the differences between the halo merger trees extracted from nine different tree building algorithms. When the evolutionary histories of individual galaxies are compared, the effect of different halo merger trees is apparent. Even though we selected the same haloes at $z=0$ from a common halo catalogue, the galaxies in them show a variety of growth histories, due to differences in the halo merger trees. The mean star formation histories of modeled galaxies slightly vary between trees, but the peak values of SFR in the most massive group of galaxies can be different by almost a factor of three. 
This difference is especially remarkable at high redshifts, where haloes are smaller and less clustered than at $z=0$. The number density of satellite galaxies is markedly different between the trees, while that of the main haloes rapidly converged with time. This naturally gives rise to different galaxy merger rates and stellar accretion fractions. In general, the trees with lower satellite number densities have lower merger rates and fractions of accreted stellar mass. Overall, tree building algorithms can result in different galaxy formation and assembly times. The time intervals between them, however, follow the trend in which more massive galaxies are formed earlier and assembled later than their smaller brethren.

Tree building algorithms that do not utilise particle identification information but rather rely upon spatial matching of haloes between snapshots (such as \jmerger) should be used with caution. 
It is difficult for such methods to accurately trace haloes back in time, which results in a truncation of their halo history.
Thus the age and evolutionary history of an object are severely curtailed. This leads to a poor representation of the evolution of the galaxy population. 

Many of the tree building algorithms tested here give similar results, but as has been shown previously this is because they have very similar underlying algorithms. These methods all make either none or limited attempts to correct for haloes missing between two snapshots. This agreement in no way makes them the best choice and we make no such inference.

Some of our tree building algorithms, \hbt, \consistenttree\ and to some extent \dtree\ and \sublink\ make attempts to correct for dropouts in the underlying halo catalogue either by rewriting the catalogue itself in the former two cases or patching up a gap in the latter two cases. While this can demonstrably increase the main branch length and therefore the length of time during which a galaxy can be seen, it does not universally improve everything.
For instance, ``improving'' the halo catalogue can lead to the removal of legitimate satellite haloes. 

We have deliberately employed only a single SAM model to illustrate the differences between the tree building algorithms. As such it is the scatter between the models which is the important characteristic here. They indicate that for any particular SAM, coupled to a simulation from which haloes were extracted using a standard procedure, a range of physical properties (such as the global star formation rate) could be obtained simply by varying the algorithm used to link the set of dark matter halo catalogues together into a tree structure. The difference between the bulk of our basic (and similar) algorithms (\mergertree, \ysamtm, \velociraptor, \treemaker) and a full tracking algorithm (\hbt) indicates how far such results would be expected to change if the tree building and underlying halo finding were to some extent completed in ideal manner. 

As summarized above, models based on the various merger trees show {\em noticeable differences} in the model galaxy properties. However, they can still be calibrated to fit some basic empirical distributions by adopting physically reasonable parameters. We are tempted to conclude that the current level of variety in merger tree building algorithms, albeit seemingly large, {\em does not result} in significant differences in synthetic galaxy properties that is significant when compared with the uncertainties in the empirical data and in other input physics in the modeling procedure. One notable finding, however, is that the use of in-principle ``superior'' algorithms does not lead to better reproduction of the empirical data, at least in our tests. More robust testing will be possible when empirical data and other input physics are better constrained.

\section*{acknowledgments}
SKY acknowledges support from the National Research Foundation of Korea(Doyak 2014003730). Numerical simulations were performed using the KISTI supercomputer under the programme of KSC-2012-C3-10. This study was performed under the DRC collaboration between Yonsei University and the Korea Astronomy and Space Science Institute.

AK is supported by the {\it Ministerio de Econom\'ia y Competitividad} (MINECO) in Spain through grant AYA2012-31101 as well as the Consolider-Ingenio 2010 Programme of the {\it Spanish Ministerio de Ciencia e Innovaci\'on} (MICINN) under grant MultiDark CSD2009-00064. He also acknowledges support from the {\it Australian Research Council} (ARC) grants DP130100117 and DP140100198. He further thanks Red House Painters for japanese to english.

PJE is supported by the SSimPL programme and the Sydney Institute for Astronomy (SIfA).

PAT acknowledges support from the Science and Technology Facilities Council (grant numbers ST/I000976/1 \& ST/L000652/1).

PSB is funded by a Giacconi Fellowship through the Space Telescope Science Institute, which is operated by the Association of Universities for Research in Astronomy, Incorporated, under NASA contract NAS5-26555.

The authors contributed in the following ways to this paper: JL \& SKY designed the comparison and performed the analysis. JL is a PhD student supervised by SKY. They along with PJE \& FRP wrote the paper. PAT, CS, AK, FRP \& AS organized the Sussing Merger Trees workshop at which this work was initiated. The other authors (as listed in section 2) provided results via their respective tree building algorithms. All authors also helped proof-read the paper.

\end{document}